\documentclass[conference, a4paper]{IEEEtran}
\IEEEoverridecommandlockouts
\usepackage{cite}
\usepackage{algorithm}
\usepackage{algorithmic}
\usepackage{amsmath,amssymb,amsfonts,amsthm}

\usepackage{algorithmic}
\usepackage{graphicx}
\usepackage{textcomp}
\usepackage{xcolor}
\usepackage{caption} 
\usepackage{stfloats} 
\usepackage{booktabs}
\usepackage{float}

\def\BibTeX{{\rm B\kern-.05em{\sc i\kern-.025em b}\kern-.08em     
		\kern-.1667em\lower.7ex\hbox{E}\kern-.125emX}}

\begin{document}
	
\title{D-CAST: Distributed Consensus Switch in Wireless Trustworthy Autonomous System}

\author{\IEEEauthorblockN{Dachao Yu}
\IEEEauthorblockA{\textit{School of}\\ \textit{Information and Control Engineering,} \\
\textit{Qingdao University of Technology}\\
dachaoyu1994@gmail.com}
\and

\IEEEauthorblockN{Jiayuan Ma}
\IEEEauthorblockA{\textit{College of}\\ \textit{ Electronic and Information Engineering} \\
\textit{Tongji University}\\
AndyMa5@outlook.com}

\and

\IEEEauthorblockN{Hao Xu}
\IEEEauthorblockA{\textit{College of}\\ \textit{ Electronic and Information Engineering} \\
\textit{Tongji University}\\
Hxu@tongji.edu.cn}
}

\maketitle
		
\begin{abstract}
The protocols of distributed consensus normally aim to tolerate different types of faults including crash faults and byzantine faults that occur in the distributed systems. However, the dynamic network topology and stochastic wireless channels may cause the same trustworthy system to suffer both crash fault and byzantine fault. This article proposes the concept of a distributed consensus autonomous switch mechanism in trustworthy autonomous systems (D-CAST) to reach the different fault tolerance requirements of the dynamic nodes and discusses the challenges of D-CAST while it is implemented in the wireless trustworthy system. 
\end{abstract}


\section{Introduction}\label{section1}
The concept of Trustworthy Autonomous Systems (TAS) has gained significant prominence since it is applied to wireless networks. These systems, characterized by their ability to operate independently and make critical decisions without human intervention, are becoming increasingly integral in various fields such as autonomous driving, industrial internet of things, and metaverse. However, the dynamic and unpredictable nature of wireless environments poses unique challenges to TAS, especially from the perspective of maintaining consistency and reliability in critical decision-making. 

Distributed consensus, which has been widely used in distributed ledger technology (DLT), can solve the challenges caused by wireless communication in TAS \cite{xu2023decontroller}. The distributed consensus aims to ensure all normal nodes in a network can achieve agreements on unified states even if the network encounters faulty progress or malicious attack \cite{xiao2020survey}. In TAS, distributed consensus can work as an internal algorithm that regulates critical decision-making and processing based on the information collected by nodes, which allows participant nodes to transmit and receive commands to change their states by following specific fault-tolerant protocols. Crash fault tolerance (CFT) and Byzantine fault tolerance (BFT) protocols are two types of distributed consensus protocols. CFT protocols like Raft \cite{ongaro2014search} and Paxos \cite{lamport2001paxos} are implemented to manage reliable state duplication and prevent system breakdown from node crash failure, whereas BFT protocols like Practical Byzantine Fault Tolerance (PBFT) \cite{castro1999practical} and Hotstuff BFT \cite{yin2019hotstuff} intends to help with the network against potential malicious attacks and arbitrary behaviors.

Based on their characteristics in fault tolerance, the BFT and CFT consensus protocols have been separately deployed in traditional distributed systems with constant network conditions but different scenarios of fault tolerance requirement \cite{klaine2023privacy} \cite{xu2023web}.  In wireless TAS, however, the dynamic node entry/exit, stochastic fluctuations of wireless channels, and the emergence of novel attack mode can cause both crash fault and byzantine fault to occur in the same network, which means the network with the configuration of only CFT or BFT protocol cannot support the reliable critical decision making or processing within it. 

In response, people are exploring adaptive schemes capable of dynamically adjusting the fault tolerance strategies to suit the prevailing conditions. The essential approach lies in the idea of autonomous switchable consensus, which allows TAS to adapt its consensus algorithms in response to changing network conditions and operational demands. This adaptability is crucial in wireless environments where factors such as signal interference, variable latency, and bandwidth constraints can significantly impact system performance. By employing a switchable consensus mechanism, TAS can seamlessly transition between different consensus algorithms, such as from a less resource-intensive algorithm during times of stable connectivity to a more robust algorithm in the face of network volatility.

This article delves into the characteristics of distributed consensus autonomous switch mechanism, exploring why it can be integrated into wireless TAS frameworks while ensuring security and scalability. Furthermore, we address the challenges and relevant solutions of distributed consensus autonomous switch mechanism in trustworthy autonomous systems (D-CAST), which includes the timing of switching to appropriate consensus algorithms based on specific conditions of the network and outline the process of leader change and state synchronization across the system during a consensus switch to maintain coherence and reliability.    

\section{Review of current applications with switchable consensus}\label{section12}

Switchable consensus mechanisms are normally implemented in DLT-based applications to provide adaptability, security, and scalability by allowing distributed systems to choose different distributed consensus protocols based on varying network conditions or performance requirements while maintaining security through robust cryptographic techniques and fault tolerance.  The switchable consensus mechanisms rely on specific governance mechanisms by distributing decision-making power among validators or stakeholders to prevent centralization. Common switchable consensus mechanisms, which can be sorted as module switch and inter-chain switch, have been applied in different DLT-based systems. This section introduces these two types of switchable consensus schemes with the example of existing developed DLT-based applications.

\subsection{Hyperledger Fabric}

Unlike public blockchains like Bitcoin or Ethereum, Hyperledger Fabric is an open-source permissioned blockchain framework, which means the parties that join the network are known and verified entities\cite{androulaki2018hyperledger}. It is part of the Hyperledger suite hosted by the Linux system, designed for use in enterprise contexts. Fabric is known for its high degrees of confidentiality, resiliency, flexibility, and scalability. The framework of Hyperledger Fabric is highly modular, which enables users to plug in their preferred components like consensus and membership services \cite{cachin2016architecture}. Fabric proposes the concept of channels to enable a group of participants to create a separate ledger of transactions, which ensures that only relevant parties can view certain transactions. 

In Hyperledger Fabric, the consensus can be switched or configured by the network's requirement, which is supported by the modular architecture of Fabric. Users change the type of consensus protocols by updating the network configuration, which involves modifying the channel configuration and getting the requisite number of signatures from the network members according to the channel's policy. The new consensus mechanism then becomes part of the channel's configuration and dictates how transactions are ordered and validated in Hyperledger Fabric.
     
\subsection{Cosmos}
Cosmos is a decentralized network within independent parallel blockchains, which can be powered by different consensus algorithms \cite{kwon2019cosmos}. The framework of Cosmos aims to enable different blockchains to communicate with each other and switch the consensus through inter-blockchain communication (IBC) protocol \cite{amoordon2019presenting}. It allows multiple blockchain systems to transfer tokens and data among nodes while maintaining their consistency of state.

Cosmos mainly uses Tendermint, which is a proof-of-stake (PoS) algorithm known for its high efficiency, speed, and robustness \cite{buchman2016tendermint}. Tendermint separates the blockchain's application layer, which can process transactions and update the state, from the consensus layer, making it simpler for developers to design blockchain systems in Cosmos. Cosmos also offers the Cosmos Software Development Kit (SDK), a framework for building blockchain applications with Tendermint. This modular framework allows developers to create customizable, interoperable blockchains with different consensus to improve the scalability of the overall system. If the Cosmos network intends to switch its consensus mechanism, it can cause a complicated problem because the new consensus algorithm may not be compatible with the existing chain's data and state. 

While Cosmos uses Tendermint Core as its standard consensus protocol, the architecture allows different blockchains within the network to operate with their consensus mechanisms. This means that within the Cosmos ecosystem, various blockchains can use different consensus algorithms, but they can still communicate with each other through IBC, provided they adhere to the IBC standards.

\section{Consensus protocols for autonomous switch}\label{section2}
The frequent critical decision-making in wireless trustworthy autonomous systems requires an instantaneous consensus switch mechanism instead of manual consensus switch schemes in Hyperledger Fabric or Cosmos. This new consensus switch mechanism needs to have the following features or capabilities:

\begin{itemize}
\item Besides the fault caused by crash nodes or malicious attacks, wireless autonomous systems may suffer from failed or delayed connection caused by dynamic channel conditions and limited communication resources. These failures are more stochastic than ordinary crash or byzantine faults in wired distributed networks. Switchable consensus mechanisms should allow the network to adapt to these changes and maintain optimal performance.  
\end{itemize}  

\begin{itemize}
\item As the network grows in size, the initial consensus mechanism might not scale efficiently. The ability to switch to the consensus mechanism with higher scalability can be crucial for maintaining the performance and reliability of the network as it expands. A mechanism that is secure for a small, closely-knit network might become vulnerable as the network grows or becomes more open.  The ability to switch to a more secure consensus mechanism is vital for maintaining the integrity of the network.
\end{itemize}  

\begin{itemize}
\item In wireless networks, especially those involving mobile or IoT devices, energy consumption is a critical concern.  Some consensus mechanisms consume more energy than others. The ability to switch to a more energy-efficient consensus mechanism can prolong the battery life of devices and reduce overall energy consumption. CFT and BFT consensus mechanisms have different thresholds of fault tolerance to the node and link failures. The ability to switch consensus mechanisms can enhance the network's robustness, especially in environments where node failures are common.
\end{itemize}  

\begin{figure}[h]   
	\centerline{\includegraphics[scale = 0.2]{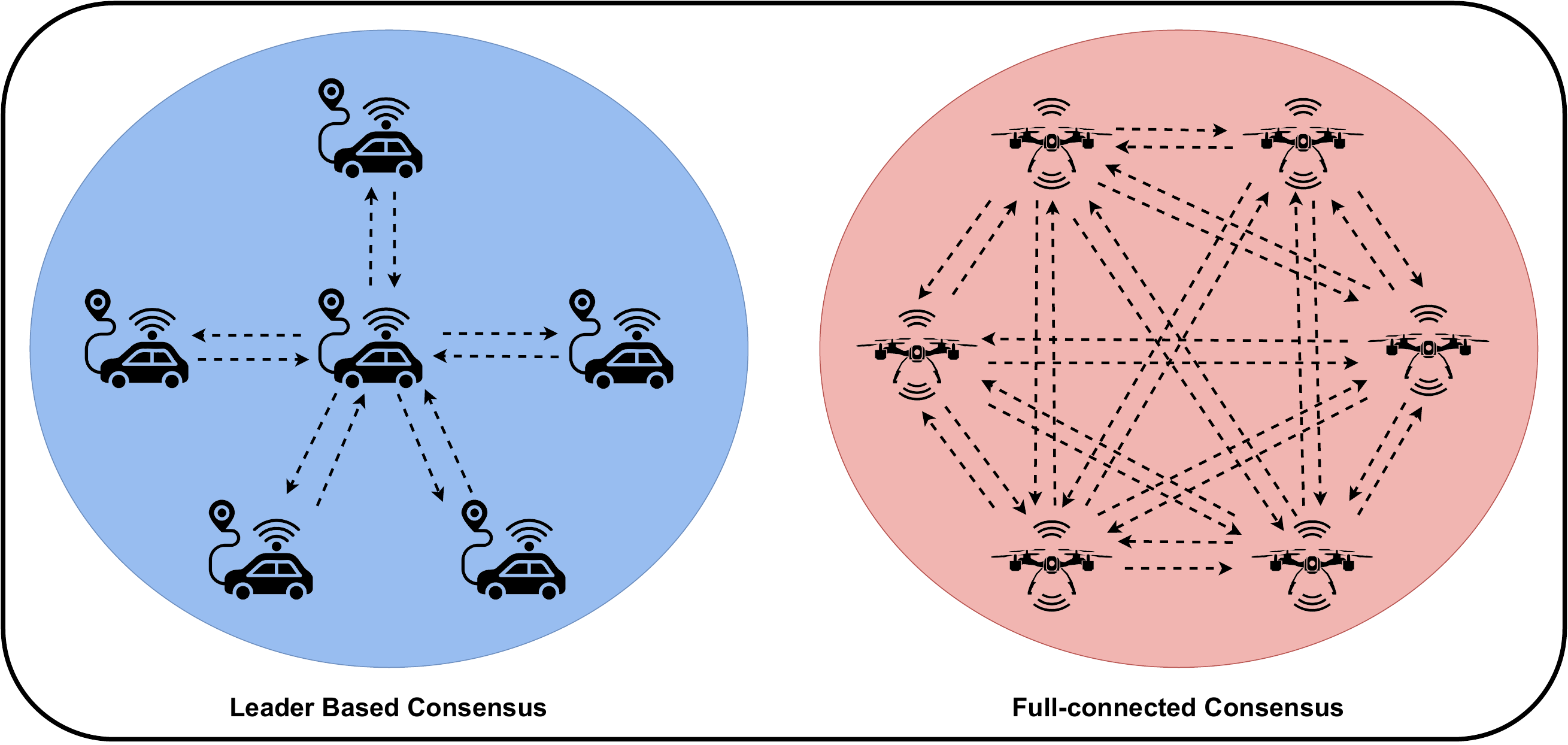}}
	\caption{Different communication topology of distributed consensus}
	\label{fig0}
\end{figure}

The communication complexity, which can be different in the protocols of consensus with various communication topologies, has a tremendous influence on the performance of wireless distributed consensus. Fig. \ref{fig0} shows two types of communication topology in the wireless distributed consensus: leader-based and full-connected. The leader-based consensus represents all nodes in a consensus-enabled network that only need to communicate with a temporary leader node that is pre-voted or selected randomly from the network. The full-connected consensus requires all nodes to communicate with each other to achieve log replication or state synchronization. The full-connected consensus has a much higher communication complexity O($N^2$) than leader-based consensus O(N), which requires more communication resources to reach the target performance when it is implemented in a wireless autonomous system.

Therefore, the leader-based consensus can be more practical to work in wireless TAS than the full-connected consensus if the TAS has limited communication resources or terrible channel conditions. Two leader-based consensus protocols are selected in the proposed autonomous consensus switch mechanism and their procedure and characteristics are introduced in this section.

\subsection{Leader-based CFT consensus: Raft}

Raft, as a typical CFT consensus algorithm, is generally implemented in a private, trustworthy, distributed system to oppose the breakdown of replicas \cite{ongaro2014search}. The main procedure of Raft has two stages, leader election and log replication. The followers need to vote for the first candidate that successfully sends the voting request to them, which means the leader candidate with the most reliable wireless connections and lowest latency is most likely to be chosen as a leader.

In the log replication, the leader needs to pack the commands in log entries and replicate the entries to all followers ceaselessly through downlink transmission. Depending on the successful reception of log messages, the followers need to reply confirmation packets to the leader through uplink unicast and start to execute the confirmed commands.  A successful log replication represents that more than 50\% overall followers have received the log entries from the leader and sent the confirmation back to the leader successfully within the time interval of two neighbor heartbeat commands. The performance of Raft is influenced by the reliability of link transmission when it is implemented in wireless networks \cite{yu2020low}. 

\subsection{Leader-based BFT consensus: Hotstuff}
Hotstuff, as a BFT state machine replication protocol, aims to ensure that non-faulty replicas agree on the order of execution for client-initiated service commands in the distributed system with $N\geq 3f+1$ replicas, despite the efforts of $f$ Byzantine replicas\cite{yin2019hotstuff}. The Libra blockchain has implemented the Hotstuff BFT as the consensus protocol in its project because of the responsiveness and linearity in this protocol \cite{baudet2019state}. The view, as a round of consensus in Hotstuff BFT, contains three phases: prepare, pre-commit, and commit. The basic Hotstuff protocol works in a succession of view numbers with monotonically increasing view numbers. Each view number has a unique dedicated leader for other replicas. The leader must collect votes from a quorum of $N-f$ replicas in three phases. The collection of $N-f$ votes to one leader is referred to as a quorum certificate (QC), which is associated with a particular node and a view number.     

In the prepare phase, the leader should collect new-view messages from replicas and process these messages to select a branch that has the highest preceding view. If over $N-f$ replicas have sent their messages to the leader successfully, a valid prepare QC can be formed through a threshold signature scheme. Then the leader broadcasts prepare QC in pre-commit messages to replicas, and replicas respond to the leader with the pre-commit vote if they receive and verify the prepare QC. Similarly, while receiving more than $N-f$ pre-commit votes, the leader combines them into a pre-commit QC and broadcasts it in commit messages to replicas. Over $N-f$ replicas should send the commit vote back to the leader for the final commit QC combination. After a successful commit QC assembly, the leader needs to send a decision message to all other replicas to notify them the consensus protocol in this view is completed, and the replicas will update their state according to the decision message. 

Compared with the high communication complexity $O(N^2)$ in the phases of other BFT consensus protocols \cite{castro1999practical}, Hotstuff has lower communication complexity, which keeps $O(N)$ in all three phases of the consensus protocol. In the scenario of wireless communication, low communication complexity can cause less interference and bandwidth cost \cite{yu2023communication}. However, because the basic Hotstuff BFT has three phases in its protocol, the communications between the leader and replicas in Hotstuff will cause longer time latency than a BFT protocol with fewer consensus phases.

According to the statement of basic Hotstuff BFT, the leader only needs to collect the vote message in the form of threshold signatures through all three phases. This feature ensures that changing the view on every prepare phase of different proposals cannot actively influence the reliability of the consensus. Therefore, the basic hotstuff protocol can be adapted to a pipeline scheme and the prepare message of the next view is binding with the pre-commit message of the former view in the same package \cite{yin2019hotstuff}. Every new view of chained Hotstuff BFT needs to start automatically after the prepare phase of the last view. This approach, which is named chain Hotstuff, aims to significantly improve the throughput of the views change in the Hotstuff BFT consensus protocol because the average latency for every view in chain Hotstuff will be shorter than the basic Hotstuff protocol if the success rate of the consensus is reasonably high \cite{yu2022centralized}.   

\section{Timing of consensus switching}\label{section3}
Although the switchable consensus can improve the performance of wireless TAS, some crucial problems should be addressed before it is implemented in practical applications. The most critical procedure of autonomous consensus switches is determining the timing of switch between Crash Fault Tolerance (CFT) and Byzantine Fault Tolerance (BFT) protocols. This section tries to discuss the solution to this challenge from the perspective of latency, the ratio of malicious nodes, and bandwidth allocation.

The success of consensus switch heavily relies on accurately identifying the optimal boundary for transitioning between these protocols. The wireless TAS must possess the capability to evaluate various factors to determine the opportune moment for protocol transition.  The primary considerations are assessing the network latency, including gauging the feasibility and effectiveness of consensus switching.

\subsection{Latency}
The ability to swiftly switch between protocols without causing disruptions or delays is paramount for maintaining the integrity and functionality of the distributed system. Time latency emerges as a dominant factor in the process of autonomous consensus switch because it directly impacts the responsiveness and liveliness of the distributed consensus mechanism. The latency of wireless distributed consensus is influenced by various factors, including the number of nodes in the network, the efficiency of wireless transmission among nodes, network congestion levels, and the computational overhead associated with protocols \cite{sakic2017response}. D-CAST mechanism should ensure the functioned consensus can reach the latency requirement of the wireless TAS or it needs to switch to the consensus protocol with less time-sensitivity.

\begin{figure*}[h]   
	\centerline{\includegraphics[scale = 0.2]{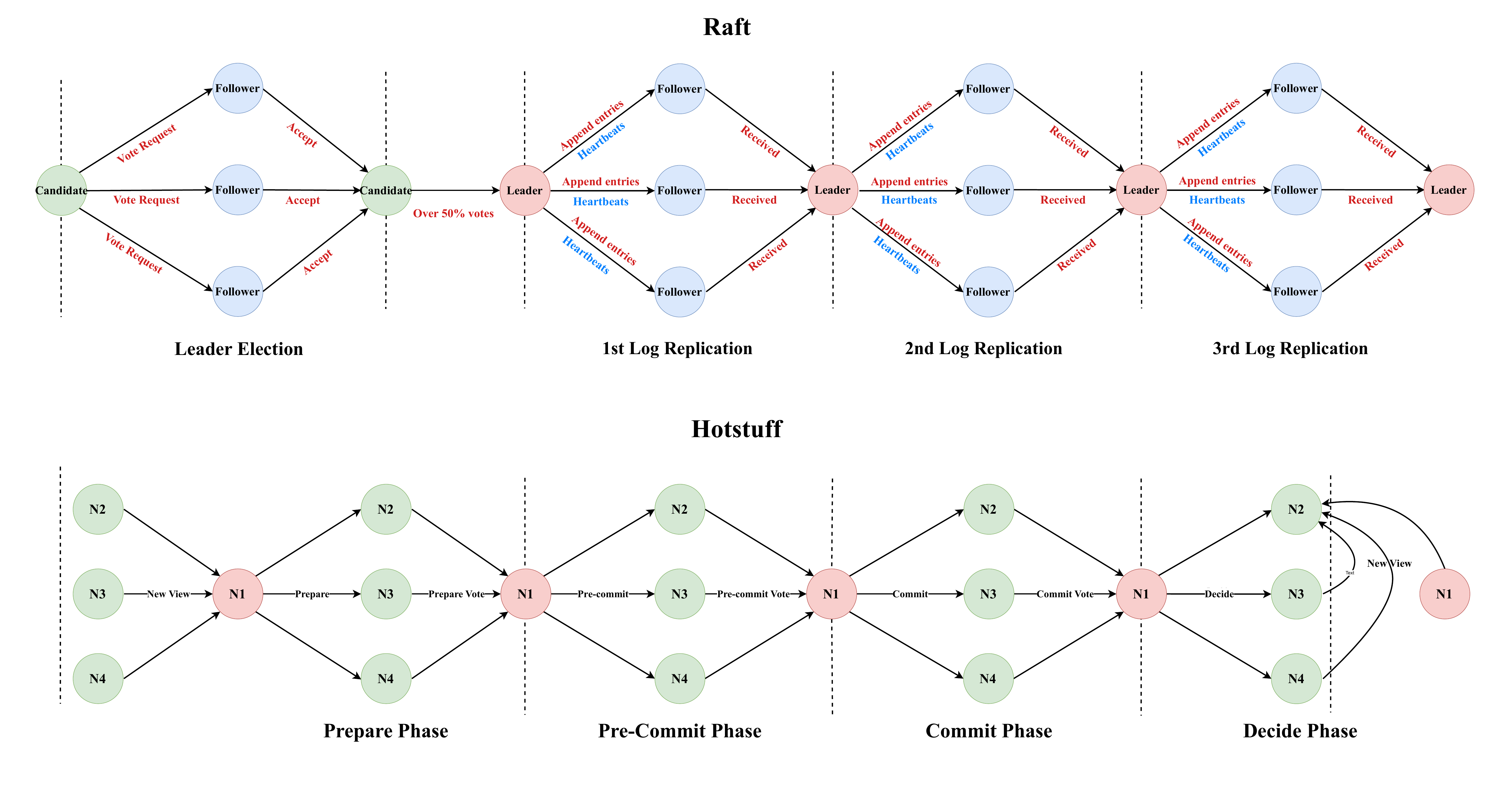}}
	\caption{The procedures of two leader-based consensus protocols}
	\label{fig10}
\end{figure*}

\subsection{Ratio of malicious node}
The dynamic switching BFT and CFT can be regarded as a hybrid approach based on observed node conditions to balance the trade-off of performance and cost. The switch between Crash Fault Tolerance (CFT) and Byzantine Fault Tolerance (BFT) consensus depends on the necessity of the system's tolerance to both crash faults and malicious behavior. In environments with a low ratio of malicious nodes, where most node failures are caused by crash, employing a CFT consensus algorithm with simplicity and efficiency might be sufficient to tolerate the failures. Conversely, in scenarios with a high ratio of malicious nodes, BFT consensus becomes necessary to maintain the reliability of critical decision-making in the system with more potential Byzantine faults \cite{9569351}. BFT algorithms provide stronger security guarantees by tolerating malicious behavior, albeit at the cost of increased communication complexity and overhead compared to CFT algorithms. 

\subsection{Bandwidth} 
In wireless autonomous systems, the decision to switch between Crash Fault Tolerance (CFT) and Byzantine Fault Tolerance (BFT) consensus is also linked to the available bandwidth allocated to channels. Bandwidth constraints significantly impact the performance of wireless distributed consensus, with CFT consensus being favored in environments with limited bandwidth because it has lower communication overhead and simplicity. Conversely, Byzantine Fault Tolerance (BFT) consensus, while imposing higher bandwidth requirements with additional message exchanges, becomes indispensable in scenarios where the risk of malicious behavior or Byzantine faults is pronounced, ensuring data integrity and system resilience against malicious nodes. The distributed consensus in TAS can dynamically transition between CFT and BFT consensus based on bandwidth allocation, enabling the system to optimize performance and the type of fault tolerance in response to fluctuations in bandwidth availability within wireless autonomous systems.

By accurately assessing the condition of network, wireless TAS can process smooth transitions between CFT and BFT protocols while maintaining the reliability and throughput of the wireless distributed network. This proactive approach to protocol management is essential for adapting to dynamic network conditions and mitigating potential risks or performance bottlenecks.

\section{Leader change and state consistency}
Besides the timing determined for consensus transitions, Fig. \ref{fig10} presents the main procedures of two leader-based consensus, which indicates that Raft and Hotstuff have different leader selection and log replication mechanisms in their protocols.
Therefore, when these two protocols are transitioned from each other, the consensus switch mechanism must be adapted to manage leader changes and maintain state consistency within the nodes of TAS.

\subsection{Leader change}
In leader-based distributed consensus, dynamic leader nodes are imperative to sustaining uninterrupted consensus progress. These leader nodes usually are determined by different mechanisms in consensus protocols. For example, the leader in Raft is elected by other followers but in Hotstuff the leader is selected by a predefined rotation mechanism. If TAS switches its consensus between these two consensus protocols, it should first expire the current leader node and use the new leader selection mechanism to choose a valid leader for further state replication.

The leader selection during the progress of transitioning from CFT to BFT should consider factors such as node reputation, performance, and cryptographic mechanisms for ensuring trustworthiness.  It ensures that the new leader is trustworthy and capable of maintaining the consensus process. Conversely, when switching from BFT to CFT, the leader selection process may become simpler, possibly involving a straightforward voting or round-robin approach \cite{andoni2019blockchain}, as the TAS no longer requires Byzantine fault tolerance.

Additionally, the autonomous consensus switch must prioritize state consistency, ensuring that most of nodes within the network maintain a coherent view of the shared state despite changes of leader node or protocol configurations. It involves implementing mechanisms to synchronize and reconcile state information among nodes, preventing inconsistencies of states that may arise from the dynamic switch between consensus protocols. 

\subsection{State consistency}
Maintaining state consistency in the TAS when switching consensus algorithms from Crash Fault Tolerance (CFT) to Byzantine Fault Tolerance (BFT) or vice versa requires careful coordination and synchronization mechanisms. A checkpointing mechanism can be implemented in it, where the state of the system is periodically saved, allowing nodes to recover to a consistent state if necessary \cite{leu1987concurrent}. In BFT consensus, strong consistency is typically guaranteed by replicated state machines, ensuring that all normal nodes agree on the order and execution of critical decisions. When it transits to CFT consensus, state consistency mechanisms may become less stringent, as CFT algorithms provide eventual consistency rather than immediate consistency. However, it's still important to ensure that all normal nodes eventually converge to the same state. Additionally, validation mechanisms should be in place to ensure that the state transition is legitimate and consistent with the rules of the new consensus algorithm \cite{ferdous2020blockchain}. It may involve cryptographic signatures or hash-based validation techniques to verify the integrity of the state transition.

\subsection{Fraud consensus switching}
The issues of leader change and state consistency indicate that the frequent consensus switch can compromise the performance of TAS.    Therefore, TAS may suffer a special attack called fraud switching from some malicious nodes, which changes the network condition significantly to increase the frequency of consensus switch and damage TAS's performance.

Some solutions can be implemented to reduce the threat of fraud switching attacks within TAS. Firstly, integrating advanced cryptographic protocols like digital signatures and robust hash functions can heighten consensus integrity, thereby raising the bar for malicious entities seeking to manipulate network conditions \cite{hafeez2023blockchain}.  Moreover, the implementation of anomaly detection algorithms is vital for real-time identification and mitigation of suspicious activities \cite{patcha2007overview}, enabling swift responses to potential fraud switching attempts. By synergizing these proactive measures, TAS may significantly fortify its resilience against fraud switching attacks, safeguarding the integrity and reliability of its operations within the network.

\section{conclusion}
This article introduces the concept of distributed consensus autonomous switch (D-CAST) in the wireless trustworthy autonomous system and discusses the challenges of the consensus switch mechanism implemented in TAS, including determining the timing of the consensus switch, leader change, and state consistency issues represented by a walk-through of consensus states. Further research is set to focus on the design of a specific consensus switch mechanism from the perspective of protocol to solve potential challenges in mission-critical wireless systems.

\bibliographystyle{ieeetr}
\bibliography{citation}

\end{document}